\documentclass[sigconf]{acmart}
\acmConference[ISSTA 2023]{ACM SIGSOFT International Symposium on Software Testing and Analysis}{17-21 July, 2023}{Seattle, USA}
\usepackage[linesnumbered, noend, lined, ruled]{algorithm2e}
\usepackage{graphicx}
\usepackage{textcomp}
\usepackage{hyperref}
\usepackage{todonotes}
\usepackage{listings}
\usepackage{multirow}
\usepackage{tikz}
\usepackage{lmodern}  
\usepackage{amsmath}  
\usepackage{xcolor}   
\usepackage{listings}
\usetikzlibrary{calc,tikzmark}
\usetikzmarklibrary{listings}
\usetikzlibrary{backgrounds}

\AtBeginDocument{%
  \providecommand\BibTeX{{%
    \normalfont B\kern-0.5em{\scshape i\kern-0.25em b}\kern-0.8em\TeX}}}

\newcommand{\codelink}{\url{https://github.com/WarningCas/WarningCascading_Data}\xspace}

\newcommand{\ben}[1]{\textcolor{black}{#1}}

\newcommand{\mycomment}[1]{}
\begin{document}

\title{A Study of Static Warning Cascading Tools (Experience Paper)}

\author{Xiuyuan Guo}
\affiliation{%
  \institution{Iowa State University}
}
\email{xiuyuang@iastate.edu}

\author{Ashwin Kallingal Joshy}
\affiliation{%
  \institution{Iowa State University}
}
\email{ashwinkj@iastate.edu}

\author{Benjamin Steenhoek}
\affiliation{%
  \institution{Iowa State University}
}
\email{benjis@iastate.edu}

\author{Wei Le}
\affiliation{%
  \institution{Iowa State University}
}
\email{weile@iastate.edu,}
 
\author{Lori Flynn}
\affiliation{%
  \institution{Carnegie Mellon University}
}
\email{lflynn@cert.org}

\begin{abstract}
Static analysis is widely used for software assurance. However, static analysis tools can report an overwhelming number of warnings, many of which are false positives. Applying static analysis to a new version, a large number of warnings can be only relevant to the old version. Inspecting these warnings is a waste of time and can prevent developers from finding the new bugs in the new version. In this paper, we report the challenges of {\it cascading warnings} generated from two versions of programs. We investigated program differencing tools and extend them to perform warning cascading automatically. Specifically, we used textual based diff tool, namely {\it SCALe}, abstract syntax tree (AST) based diff tool, namely {\it GumTree}, and control flow graph (CFG) based diff tool, namely {\it Hydrogen}. We reported our experience of applying these tools and hopefully our findings can provide developers understandings of pros and cons of each approach. In our evaluation, we used 96 pairs of benchmark programs for which we know ground-truth bugs and fixes as well as 12 pairs of real-world open-source projects. Our tools and data are available at \codelink.

\end{abstract}

\maketitle

\section{Introduction}
In an agile software development setting, there is a need to deliver reliable new software releases in a rapid fashion. The big challenge is how we can only analyze and report the software quality issues related to the new version, as the issues of old versions have been addressed previously when shipping the old versions. In particular, static analysis tools, as an important software assurance technique, often generate an overwhelming number of warnings for each version of software \cite{imtiaz_challenges_2019,christakis_what_2016}. It is confusing which warnings are related to only the old code and have been reviewed in the previous versions, which warnings report new issues in the updated release, and which warnings are about the issues of fixing the old warnings. Consequently,  tremendous time and manual efforts can be wasted and not spent on the right problems of the current version.



The goal of {\it static warning cascading} (also called {\it matching} or {\it aligning} static warnings) is to help developers classify warnings into several categories: (1) the cascaded warnings report a same issue in the old and new versions (so we don't need to handle it), (2) the warnings in the old version are fixed in the new version (we can inspect them together to confirm if the fix is indeed correct), (3) the warnings are changed from the old version but the old and new warnings are related (we should inspect them together to understand the problem), and (4) the warnings only report issues in the new version (we should inspect them in the new version of software). For most of static analysis tools, a warning is reported as a line in the source code file. Thus the warning cascading problem can be reduced to map a source code line from the old version to the new version and classify the mapping to be one of the above categories.


There exist a spectrum of program differencing tools~\cite{MVICFG,Falleri2014Gumtree,scale} that can match source code lines. Some representative categories include {\it textual based diff}, {\it syntax based diff} and {\it control flow based diff}. Textual based diff is typically performed on two source code files using the {\it longest common sequence} algorithm like the one implemented in the {\it Unix diff} tool. Syntactic based diff tools like  GumTree~\cite{Falleri2014Gumtree} are performed on the {\it abstract syntax trees (ASTs)}. It compares ASTs from two versions of a source code file and determines if the AST nodes in the two versions should be matched. The control flow based diff uses a representation of {\it multiple version interprocedural control flow graphs (MVICFG)}~\cite{MVICFG}. The MVICFG  is a union of {\it Interprocedural Control Flow Graphs (ICFGs)} for a set of program versions. The common nodes and edges in versions are represented only once and each edge is marked with the versions it belongs to.

In this paper, we conducted a study of the three representative program differencing techniques for cascading static warnings. Our goal is to evaluate the pros and cons of each tool and report which tools are the most {\it useful} and {\it successful} for cascading warnings. By useful and successful, we mean when a bug in a program is fixed in the new version, the tool is able to report the warning of the bug does not cascade to the one in the fixed version; when a bug in a program still exists after adding changes, the tool is able to report that the warning in the buggy version is cascaded to the new version. They are the same or related warnings. 

Specifically, we used an existing texual based warning cascading tool  {\it SCALe}~\footnote{https://github.com/cmu-sei/SCALe/tree/scaife-scale} developed and  used at CERT~\footnote{https://www.sei.cmu.edu/about/divisions/cert/}; we also designed and implemented warning cascading routines on top of other two open source tools, GumTree and Hydrogen. We applied the three tools for static analysis warnings generated for two program versions. To compare different tools' behavior when apply warning cascading, we first constructed studies on the 96 pairs of benchmark programs where the ground truth are known. We then collected 12 pairs of real-world open source projects and investigate the use of the three tools in practice.

Our results show that Hydrogen has a slight advantage compared to other two tools for  the ground truth benchmarks. When used for real-world programs such as {\tt find}, {\tt grep}, {\tt make} and {\tt coreutils},  Hydrogen is more successful for cascading  same "bugs" across versions compared to two other tools, where SCALe shows more advantages in cascading the cases where the warnings for the first version are fixed in the second versions. We sampled a set of our results and reported the analysis on these examples (see \S4 for details). Our experience and findings can provide developers knowledge on warning cascading as well as a more general problem of programming differencing.

In summary, this paper made the following contributions:

\begin{enumerate}
    \item We reported practical challenges of cascading warnings across program versions and proposed what is considered as a successful warning cascading (\S 2);
    \item We used and extended three types of programming differencing tools to perform warning cascading (\S 3);
    \item We designed and performed comprehensive empirical studies to compare the three types of approaches to discover whether, when and why each type of the tools work best for warning alignment (\S 4); and 
    \item We open source our tools and datasets at \codelink. 
\end{enumerate}

\section{Motivation and Challenges}

Warning cascading is challenging because when programs are updated in the new versions, the function and variable names may change, and the line numbers of the same statements are also likely changed. Directly performing string matching for the output from static analysis tools cannot work because of the change of context in the newer version. In this section, we provide some examples to explain the challenges of warning cascading, and we also more precisely define what it means by a useful and successful cascading.

\subsection{Challenging examples}
{\tt find.71f10368} has a bug of "crashing in some locales for find -printf '\%AX'"  and a newer version of {\tt find} added a fix for this bug and also included many additional new changes. If we run static analysis tool, we will get 1600+ warnings for each version. Without a proper warning cascading tool, we cannot easily find which warnings in the previous version are changed in the new version, and determine whether the fix is successful and whether there are more new issues introduced in the fix and other newly added code.



Warning cascading is challenging for several aspects. First, there are many identical warnings between the two versions; however, the same warnings across versions may be reported as different locations of the same files due to the new code added or old code deleted, and there can be changes of variable/function names, e.g., via refactoring, which do not affect the warning semantics. Second, there are often dead code in the project, e.g., {\tt corebench}~\footnote{https://www.comp.nus.edu.sg/~release/corebench/} have \texttt{gnulib-tests} folders within the projects which did not affect program behaviors. But static analysis scans all the code to output the warnings. Developers have to filter out those warnings irrelevant to the newly developed code. Such dead code can be project specific and hard to exclude, and thus increase the overhead of warning cascading.

Here, we further show some real-world examples discovered in our study. In the first case, when many new lines are added before the target line, the text diff tools cannot match the warnings. See Figure~\ref{img:chg1}. In the second case, a line added in the new version (green at line 5 in Figure~\ref{img:chg2}) is the same as the target line (blue at line 7 in Figure~\ref{img:chg2}). The diff tools can be confused and mistakenly match newly added line~5 with {\tt i++} in the old version instead of line~7. In the third case, there are non-semantic changes, e.g., changing function name between the two versions or adding a new comment to the target line. As an example, in Figure~\ref{img:chg3}, \texttt{fprintf} is changed to \texttt{checked\_fprintf}, and the text diff tools cannot match them. In Figure~\ref{img:chg4}, a statement at line~7 didn't change at all in the second version but an extra comment is added. These cases can challenge the warning cascading tools.


    
    
    

    

\begin{figure}
    \centering
    \begin{lstlisting}[language=C,
    numbers=right,
    frame=none,
    xleftmargin=0.5cm,
    xrightmargin=1.5cm,
    basicstyle=\footnotesize\ttfamily]
    
static char **
construct_command_argv_internal
(char *line, char **restp, char *shell,
char *shellflags, char *ifs, int flags,
char **batch_filename_ptr)
{
    ...
    
    <@\textcolor{green}{+ if(one\_shell)}@>
    <@\textcolor{green}{+ \{}@>
    <@\textcolor{green}{+ \#if defined \_\_MSDOS\_\_ || defined (\_\_EMX\_\_)}@>
    <@\textcolor{green}{+ if (unixy\_shell)}@>
    <@\textcolor{green}{+ \#else}@>
    <@\textcolor{green}{+ if (is\_bourne\_compatible\_shell(shell))}@>
    <@\textcolor{green}{+ \#endif}@>
    <@\textcolor{green}{... // more lines are added here}@>
    
    ...
    <@\textcolor{blue}{command\_ptr = ap;}@>
}
    \end{lstlisting}
    \caption{Challenge: many lines (lines~9--17) are added before the target line (line~25)}
    \label{img:chg1}
\end{figure}

\begin{figure}
    \centering
    \begin{lstlisting}[language=C,
    numbers=right,
    frame=none,
    xleftmargin=0.5cm,
    xrightmargin=1.5cm,
    basicstyle=\footnotesize\ttfamily,]
static bool
record_exec_dir (struct exec_val *execp)
{
    ...
    <@\textcolor{green}{+ i++ // newly added line in version2 could possibly match two version1 }@>
    function1()
    <@\textcolor{blue}{i++ // shared line between version1 and version2 }@>
    ...
}
    \end{lstlisting}
    \caption{Challenge: newly add a line the same as the target line}
    \label{img:chg2}
\end{figure}

\begin{figure}
    \centering
    \begin{lstlisting}[language=C,
    frame=none,
    basicstyle=\footnotesize\ttfamily,
    escapeinside={(*@}{@*)}]
static reg_errcode_t
internal_function
re_string_reconstruct (re_string_t *pstr, Idx idx, int eflags)
{
    ...
    case KIND_FORMAT:
        switch (segment->format_char[0])
    case 'a':
        (*@\textcolor{red}{- fprintf (fp,segment->text , ctime\_format (get\_stat\_atime(stat\_buf)));}@*)
        (*@\textcolor{green}{+ checked\_fprintf (fp,segment->text , ctime\_format (get\_stat\_atime(stat\_buf)));}@*)
        break;
    case 'b':
        (*@\textcolor{red}{- fprintf(fp,segment->txt, human\_readable ((unintmax\_t) ST\_NBLOCKS (*stat\_buf)));}@*)
        (*@\textcolor{green}{+ checked\_fprintf(fp,segment->txt, human\_readable ((unintmax\_t) ST\_NBLOCKS (*stat\_buf)));}@*)
        break;
    ...
}
    \end{lstlisting}
    \caption{Challenge: new name is applied during refactoring}
    \label{img:chg3}
\end{figure}

\begin{figure}
    \centering
    \begin{lstlisting}[language=C,
    numbers=right,
    frame=none,
    xleftmargin=0.5cm,
    xrightmargin=1.5cm,
    basicstyle=\footnotesize\ttfamily,]
static bool
record_exec_dir (struct exec_val *execp)
{

    ...
    i++; 
    <@\textcolor{green}{+ i++; //extra comment}@>
    ...    
}
    \end{lstlisting}
    \caption{Challenge: add a comment}
    \label{img:chg4}
\end{figure}

\subsection{What is a useful and successful warning cascading?} 
In the problem of warning cascading,  the tool takes static warnings generated from one version and determines if there is a {\it match} for the warnings in another version.  We consider a warning is successfully cascaded for the following two cases. First, the cascading tool reports the two warnings as {\it same "bugs"} if both versions contain the same "bug" located at the target line (we say the "bugs" in two versions are the same if the sequences of root cause statements along the paths are semantically equivalent). In this case, the warnings have been reviewed in the old version, and developers do not need to further investigate these warnings. Here, "bug" is not confirmed but is the output warning from static analysis tools.  "bug" can be false positives, and we can match them if the changes newly added do not affect the semantics of the warnings.

A variant of the first case is that the root cause statements of the "bug" are not exactly the same but have some changes----for a successful cascading, the tool should report the two warnings as {\it relevant "bugs"}.  So developers can inspect the two warnings together for diagnosis. 

In the second case, one version contains the "bug" at the target line and the other version added a fix for the "bug". There is no longer warning reported for this "bug" in the second version. Here, a successful cascading should report {\it "bug" fix}.  This case includes a special situation, where the buggy code is deleted in the second version. The warning cascading in this case is useful especially when the second version aims to fix the bugs in the first version. Warning cascading is able to help determine if the issues in the first version likely are addressed in the new versions, and what are the new issues added in the new version.


If the cascading tools fail to match the same "bugs" (in the first case) or match any "bug" in one version with irrelevant "bugs" in another version (in both first and second cases), we consider such cascading as unsuccessful. In our evaluation, we used benchmarks that are known with ground truth bugs and fixes to evaluate such metrics. For the real-world benchmarks where there is no ground truth, we performed manual inspection to determine if the warning cascading is successful (details see \S 4).

\section{Three techniques of Cascading Warning}
In this paper, we used three different types of program differencing tools, namely {\it textual based diff}, {\it AST based diff} and {\it CFG based diff}, for warning cascading. Specifically, {\it SCALe} is a tool developed by CERT and used {\tt Unix diff} for cascading warnings. {\it GumTree} is a  syntactic differencing tool based on abstract syntax trees (ASTs). {\it Hydrogen} compares programs based on control flow graphs integrated in MVICFG. We extended GumTree and Hydrogen for warning cascading. 

We compare the output of these tools to understand the pros and cons of these techniques. We hope our findings can help developers better select warning cascading tools and more efficiently improve their code quality in the continuous integration. In the following, we provide some technical details of the three tools.



\subsection{Textual based diff tool: SCALe}
SCALe~\cite{scale} takes warnings reported by multiple static analysis tools and first group the warnings of the same issue (reported by different static analysis tools) into one warning. It then applies the textual {\it diff} tool~\footnote{\url{https://www.gnu.org/software/diffutils}} to maps lines between versions and assess whether the source code line associated with a warning has undergone modification in the updated version. 



To apply SCALe, we deployed docker-based virtual machine~\footnote{\url{https://github.com/cmu-sei/SCALe/tree/scaife-scale}}. For each program analyzed, the static analysis tools were run, and the output was formatted in a way that SCALe could process and understand. These output files were uploaded into SCALe via their web-based interface. The warnings of the second version of the program are added in the same way. To compare the different versions of the program, we leveraged the browser interface's built-in functions and employed the {\it diff} tool for cascading the differences. This was executed internally by the browser's backend, which allowed for the comparison of the program and performed warning cascading on the program versions. After that, we can find the output of it on the GUI page, and the information contains the {\it verdict} value of each warning.

If the warning's verdict value is {\tt true}, that means there is no adjudication on the previous version for the line corresponding to the warning, and there is matched warning in the first version. If such a value is false, that means there is a change in the previous version. The warning in the second version is not matched. We review any warnings present in the first version but not matched as {\it same "bug"} in the second version and label them as {\it "bug" fix} cascading. We implemented a script to automate the process of generating warnings from multiple static analysis tools, uploading the results to SCALe, and cascading warnings between two versions of a program. 

\subsection{AST based diff tool: GumTree}
GumTree~\cite{Falleri2014Gumtree} is a syntactic differencing tool that operates based on the Abstract Syntax Tree (AST). Unlike SCALe, which uses the diff tool to compare file versions on a text line level, GumTree parses each file version into an AST representation and directly matches the nodes in the two AST versions.
By utilizing the AST representation, GumTree is able to bypass the influence of minor changes that may surround warnings, such as changes in spacing or refactoring of variable names, which are unlikely to affect the warnings. When the warnings in the two versions correspond to the aligned nodes identified by GumTree, we consider the warning is cascaded to the new version.

To perform syntactic warning cascading, we built a custom client interface for the code release of GumTree\footnote{\url{https://github.com/GumTreeDiff/gumtree}}.
GumTree's objective is to compute an \textit{edit script}, a sequence of edit actions made to a source file, which is short and close to the developer's intent. It follows a 2-step process. Step 1 computes mappings between similar nodes in the two ASTs; the main contribution of GumTree is to maximize the number of mapped nodes. Step 2 deduces the edit script from the AST mapping using the algorithm of Chawathe et al \cite{Chawathe1996Change}. We only used the AST mapping and did not compute the edit script, since our application only needs to match the AST nodes between two versions of a program.

Given two versions of a source code file $P_1$ and $P_2$, the GumTree parser produces their respective ASTs $T_1$ and $T_2$.
Then, GumTree computes the mapping $M_T$ between the similar nodes in $T_1$ and $T_2$.
Finally, our client checks the mapping $M_T$ to determine the set of warnings to cascade.

Algorithm \ref{alg:gumtree} defines our cascading algorithm built on top of the GumTree.
Since all warnings are placed on concrete lines in the code, we traverse only the leaf nodes of the AST. A warning is cascaded between nodes $t_1 \in T_1, t_2 \in T_2$ if and only if
there is at least one warning on the same line as both $t_1$ and $t_2$ (line 3),
$t_1$ is mapped to $t_2$ by GumTree (line 4), and
the warnings attached to $t_1$ and $t_2$ have the same CWE (Common Weaknesses Enumeration) condition (line 5).

Similar to SCALe's implementation of diff cascading, our implementation uses the results of running GumTree to cascade the warnings without modifying the GumTree AST parser and differencing algorithm. Our implementation preserves the stable and efficient implementation and adds only the little overhead which is necessary for cascading warnings.

\begin{algorithm}
\SetAlgoLined
\SetKwInOut{Input}{Input}
\SetKwInOut{Output}{Output}
\Input{AST for 2 versions $T_1, T_2$, node mapping $M_T = \{ (t_1, t_2) \}$}
\Output{$M_W = \{ (w_1, w_2) \}$ matched from $T_1$ to $T_2$}

$M_W \leftarrow \emptyset$

\For{$w_1 \in W_1$}
{
    \For {$t_1 \in leaves(T_1) | t_1.line = w_1.line$}
    {
        \If {$(t_1, t_2) \in M_T$}
        {
            
            $M_W += \{ w_2 \in W_2 | t_2.line = w_2.line \land w_1.condition = w_2.condition \}$
            
        }
    }
}

\Return $M_W$
\caption{\label{alg:gumtree}Cascade Warnings with GumTree}
\end{algorithm}

\subsection{CFG based diff tool: MVICFG}\label{sec:Algorithm}
Hydrogen~\cite{MVICFG} is a tool that differentiates programs with regards to its  control flow. It first generates the ICFG based on IR produced by LLVM~\cite{llvm}. It then combines ICFGs of each version into one via a graph union. The nodes and edges are shared across multiple versions and are marked with versions they belong to. In the end, it builds a program representation for multiple versions of ICFGs, called MVICFG, which shows different control flows and paths between  two different program versions. 

To perform warning cascading, we developed an extension to Hydrogen's original algorithm. This extended algorithm utilizes various graph traversals to detect the cascading of warnings, shown in Algorithms 2 and 3. 

Algorithm~\ref{alg:CascadeWarning} takes as input the two program versions ($V_1$ \& $V_2$) and their respective collection of static warnings ($SW_1$ \& $SW_2$). The algorithm outputs ($W_m$ \& $W_u$) as a matched warning (cascading bugs) and unmatched warning (cascading fixes) respectively.



We generate MVICFG of the two versions of the program at line 3 of Algorithm~\ref{alg:CascadeWarning}. Then we embed the warnings from both versions of $SW_1$ and $SW_2$ into the MVICFG at line 4, based on its location in the code, including the file path, file name, function name, and line number. After embedding, each warning contains meta-data that specifies the version from which it originates, the type and message associated with the warning, and its node in MVICFG. At lines~5 and 6, we iterate through all the warnings from $SW_1$. For each warning, we obtain its corresponding MVICFG node based on the metadata provided by the warning, and we then see if node $n$ is a common node shared across two versions~\cite{Le_Pattison_2014}. If so, the warnings at these locations have the possibility of being cascaded. If the node is a common node and it also contains the warning from the second version, we add it into $W_m$. Otherwise, if the node is not a common node, we put it into \textit{CheckBetween} (discussed later) function to further identify whether it can be cascaded.


\begin{algorithm}[h!tb]
  \SetAlgoVlined{}
  \SetKwInOut{Input}{Input}
  \SetKwInOut{Output}{Output}
  \SetKwProg{Fn}{\bf Function }{}{end}

  \SetKwFunction{EmbedInMVICFG}{EmbedInMVICFG}
  \SetKwFunction{InitializeMVICFG}{Initialize MVICFG}
  \SetKwFunction{GenMVICFG}{GenMVICFG}
  \SetKwFunction{CascadeWarning}{CascadeWarning}
  \SetKwFunction{GetMatchedNodesInMVICFG}{GetMatchedNodesInMVICFG}
  \SetKwFunction{GetWarningsAtNode}{GetWarningsAtNode}
  \SetKwFunction{GetMatchedWarning}{GetMatchedWarning}
  \SetKwFunction{CheckPath}{CheckPath}
  \SetKwFunction{GetDeletedNodesInMVICFG}{GetModifiedNodesInMVICIFG}
  \SetKwFunction{GetDivNodeInMVICFG}{GetDivNodeInMVICFG}
  \SetKwFunction{GetConvNodeInMVICFG}{GetConvNodeInMVICFG}
  \SetKwFunction{CheckBetween}{CheckBetween}
  \SetKwFunction{GetNodeFromWarningData}{GetNodeFromWarningData}

  \SetKwData{MVICFG}{MVICFG}
  \SetKwData{MN}{MN}
  \SetKwData{n}{n}
  \SetKwData{EW}{EW}
  \SetKwData{w}{w}
  \SetKwData{DN}{DN}
  \SetKwData{DivN}{DivN}
  \SetKwData{ConvN}{ConvN}
  \SetKwData{Stmt}{Stmt}
  \SetKwData{Start}{Start}
  \SetKwData{SegA}{SegA}
  \SetKwData{SegB}{SegB}
  
  \Input{Program versions [$V_1,V_2$],\\ Resp.\ warning sets [$SW_1,SW_2$]}
  \Output{Matched warnings [$W_m$],\\ Exclusive warnings [$W_u$]}
  Initialize $W_m,W_u$, \MVICFG\;
  \Fn{\CascadeWarning{$V_1,V_2,SW_1,SW_2$}} {
    \MVICFG$\leftarrow$ \GenMVICFG{$V_1,V_2$}\;
    \EmbedInMVICFG{$SW_1,SW_2$}\;
    \While{$SW_1 \neq \emptyset$} {
      Remove a warning \w from $SW_1$\;
      \n$\leftarrow$ \GetNodeFromWarningData{\MVICFG , \w}\;
      \If{\n.IsSharedNode \textbf{and} \n.HasWarning\_2}
      {
        Add \w to $W_m$\;
      }
    }
  }
  \caption{Cascade warnings}\label{alg:CascadeWarning}
\end{algorithm}

Since there is a possibility of the line being marked as modified due to changes surrounding it, Unix diff tool will report the line as changed. MVICFG used Unix diff tool and thus the changed lines will be represented in two different nodes. But we can recover such weakness of Unix diff tools by further checking if the (buggy) paths lead to this line in two versions are actually the same. If so, we can category the warnings as matched. See Algorithm 3 \textit{CheckBetween}.


\begin{algorithm}[h!tb]
  \SetAlgoVlined{}
  \SetKwInOut{Input}{Input}
  \SetKwInOut{Output}{Output}
  \SetKwProg{Fn}{\bf Function }{}{end}

  \SetKwFunction{EmbedInMVICFG}{EmbedInMVICFG}
  \SetKwFunction{InitializeMVICFG}{Initialize MVICFG}
  \SetKwFunction{GenMVICFG}{GenMVICFG}
  \SetKwFunction{CascadeWarning}{CascadeWarning}
  \SetKwFunction{GetMatchedNodesInMVICFG}{GetMatchedNodesInMVICFG}
  \SetKwFunction{GetWarningsAtNode}{GetWarningsAtNode}
  \SetKwFunction{GetMatchedWarning}{GetMatchedWarning}
  \SetKwFunction{CheckPath}{CheckPath}
  \SetKwFunction{GetDeletedNodesInMVICFG}{GetDeletedNodesInMVICFG}
  \SetKwFunction{GetDivNodeInMVICFG}{GetDivNodeInMVICFG}
  \SetKwFunction{CheckBetween}{CheckBetween}
  \SetKwFunction{FirstDivNodeInMVICFG}{FirstDivNodeInMVICFG}
  \SetKwFunction{FirstConvNodeInMVICFG}{FirstConvNodeInMVICFG}
  \SetKwFunction{StmtInMVICFG}{StmtInMVICFG}
  \SetKwFunction{GetDivNode}{GetDivNode}

  \SetKwData{MVICFG}{MVICFG}
  \SetKwData{MN}{MN}
  \SetKwData{n}{n}
  \SetKwData{EW}{EW}
  \SetKwData{w}{w}
  \SetKwData{DN}{DN}
  \SetKwData{DivN}{DivN}
  \SetKwData{ConvN}{ConvN}
  \SetKwData{Stmt}{Stmt}
  \SetKwData{Start}{Start}
  \SetKwData{SegA}{SegA}
  \SetKwData{SegB}{SegB}

  \Input{Warning \w at MVICFG node \n}
  \Output{Updated $W_m,W_u$}
  \Fn{\CheckBetween{\w,\n}} {
    \DivN$\leftarrow$ \FirstDivNodeInMVICFG{\n}\;
    \ConvN$\leftarrow$ \FirstConvNodeInMVICFG{\n}\;
    \Stmt$\leftarrow$ Statement at \n\;
    \MN$\leftarrow$ \StmtInMVICFG{\Stmt,\DivN,\ConvN,$V_2$}\;
    \If{\MN $\neq \emptyset$ \textbf{and} \MN.HasWarning\_2} {
      Add \w to $W_m$\;
    }
    \Else{
        Add \w to $W_u$
    }
  }
  \caption{Categorize matched warnings}\label{alg:CheckPaths}
\end{algorithm}

In Algorithm~\ref{alg:CheckPaths}, at lines 2 and 3, we traverse MVICFG to get the {\it divergent/convergent} nodes nearest to $n$. A {\it divergent} node of $n$ on the MVICFG is defined as a nearest {\it matched} node (matched across two versions) found by traversal of the predecessor edges from $n$. A {\it convergent} node of $n$ is a nearest matched node found by traversal of a successor edge of $n$. We provide an example to further clarify the two definitions. Figure \ref{fig:div_con} showed a snippet of MVICFG. Node $n_1$ is a matched node shared between two versions. From this node, there are two edges which are called {\it version branches} in the MVICFG. Nodes $n$ and $n_2$ on the left version-branch belong to version 1. Node $n$ on the right version-branch belongs version 2.  Here, node 1 is the divergent node for $n$ and $n2$, and $n_3$ is the convergent node for $n$ and $n2$. The two nodes $n$ along the two version branches have the same statements. We will use algorithm \ref{alg:CheckPaths} to mark those two as matched by leveraging the use of divergent and convergent nodes.  

\begin{figure}[ht]
  \centering
  \includegraphics[width=0.8\linewidth]{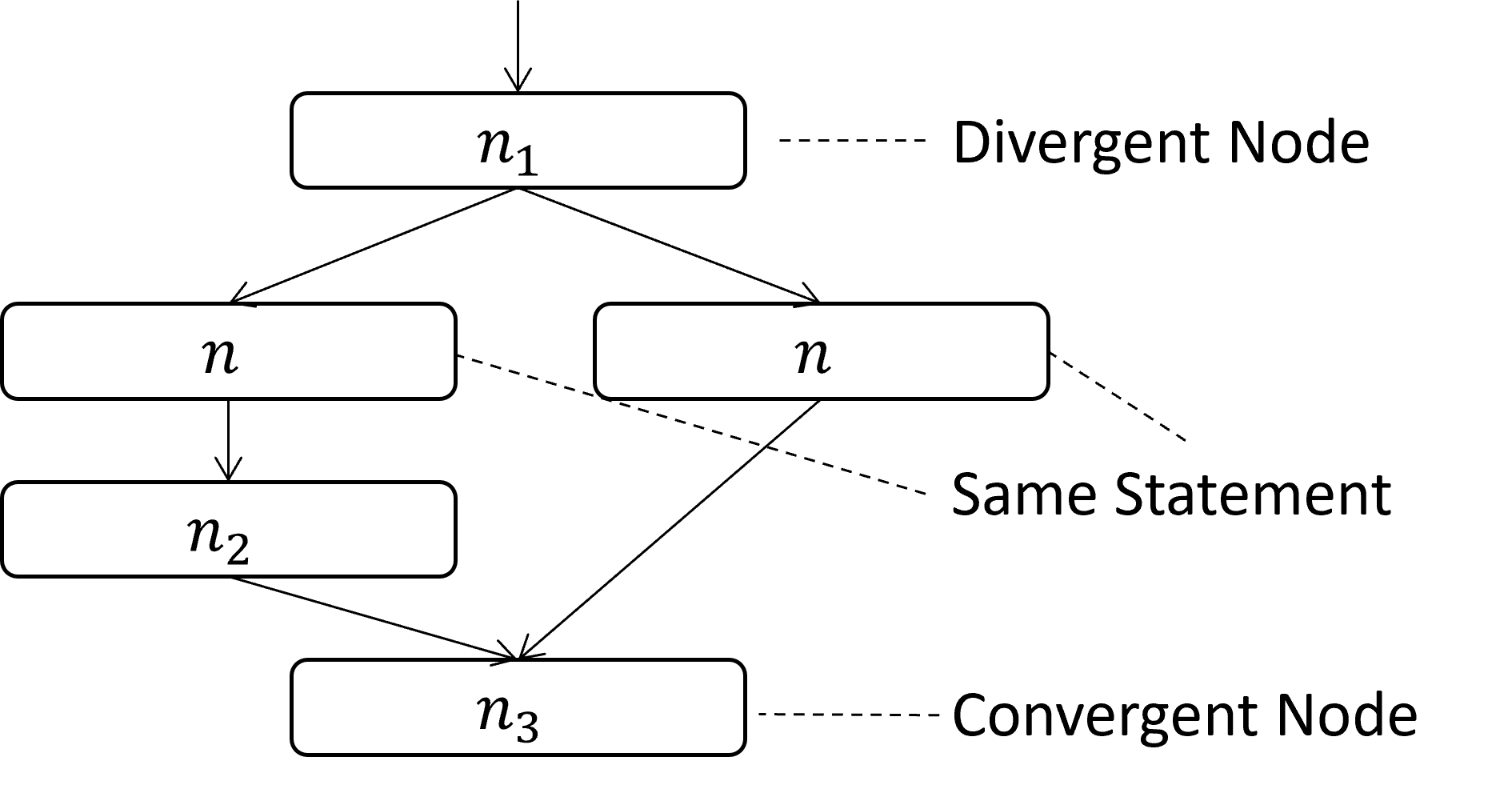}
  \caption{Divergent and Convergent Nodes on the MVICFG}
  \label{fig:div_con}
\end{figure}

Specifically, at line 2 in Algorithm 3, \texttt{FirstDivNodeInMVICFG} performs a breadth-first search backwards from \texttt{n} and returns the divergent node with the shortest path to \texttt{n}. Similarly, at line~3,
\texttt{FirstConvNodeInMVICFG} performs a breadth-first search forwards from \texttt{n} and returns the convergent node with the shortest path from \texttt{n}.
Because \texttt{n} is a modified node, there will be at least one divergent node in its ancestors and at least one convergent node in its successors.
At line 10, we extract the statement as a string and trim whitespace characters.
Then, at line 11, \texttt{StmtInMVICFG} searches all nodes between $DivN$ and $ConvN$ for a node in $V_2$ whose text exactly matches $Stmt$ after trimming whitespace.
If such a node exists and it contains a warning for the second version, we consider it as matched with \texttt{n} and categorize the warning as cascaded $w_m$; otherwise, we add it into $W_u$


\section{Evaluation}\label{sec:eval}
In the evaluation, we plan to investigate:
\begin{enumerate}
\item Which approaches perform the best for static warning cascading? 
\item When and why each approach does not perform well?
\end{enumerate}

\subsection{Experimental setup}\label{subsec:expSetup}
\paragraph{Experiments} We designed two experiments, namely a {\it ground-truth setting} and a {\it real-world setting}. In the ground-truth setting, we collected a set of buggy programs, where we know the location of the bugs in each program. Each buggy program has two variants: a {\it buggy-buggy variant} consisting of the original buggy program and a version in which a refactoring irrelevant to the bug is introduced, and a {\it buggy-fix} variant, consisting of the original buggy program and a version in which the bug is fixed. We selected the buggy programs such that at least one of our static analysis tools can correctly report the warnings for the known bugs. That way, we can count and analyze how these warnings are cascaded in its variants, and compare the cascading results with the ground truth.

In the real-world setting, we collected a set of programs consisting of real-world bugs and their fixed versions from open source projects. We then observed the cascading of static warnings from the first version to the second version. This setting helps us understand the usefulness and challenges of cascading approaches in the real-world application settings.

\paragraph{Software subject selection}
To fulfill the two experiment settings, we used C programs from two benchmarks: SARD \cite{SARD} and CoREBench~\cite{errorComplexity}. From the SARD dataset, we used ABM and Toyota as the ground truth setting. The two benchmarks consist of 96 pairs of synthetic programs, where static analysis tools are able to report warnings for the buggy version. 


CoREBench consists of a total 12 pairs of real-world projects, including \texttt{make, find, grep,} and \texttt{coreutils}.  These are open-source programs with a long contribution history of over 33k commits. Each program is documented with real-world bug reports and their corresponding fix introducing commit. The program represents a wide variety of project sizes, ranging in size from 9.4k LOC (grep) to 83.1k LOC (coreutils).


\paragraph{Static analysis tools.}
To generate the static warnings for cacading, we used five different tools: GCC~\cite{gcc}, Clang~\cite{clang}, Cppcheck~\cite{cppcheck}, Rosecheckers~\cite{rosecheckers} and CodeSonar\cite{codesonar}. These tools are currently supported by SCALe and frequently used in CERT~\cite{cert} for scanning vulnerabilities. We used SCALe first to aggregate the warnings generated from different static analysis tools into one warning. We then cascade the warning using the three tools.

\paragraph{Metrics and confirmation of the results}: For the benchmarks where we know the ground-truth bugs, we used the warnings reported at the buggy lines in the first version as subjects and determined the successfulness of warning cascading based on our criteria given in Section~2.2. 

For the real-world programs,  we sampled 12\% of total warnings from one version and manually inspected if the warnings are cascaded successfully. For each pair of programs, we report (1) the warnings of the two versions are matched as "bugs", and (2) the warnings in the first version are removed in the second versions. We then compared the results from three tools and also performed manual inspection to evaluate whether such two types of match are performed correctly by the three tools. For example, in case (1), a mistake is reported if the two warnings are not the same but paired incorrectly by a tool, or the two warnings are supposed to be matched, but one tool fails to do so; in case (2), the warning is supposed to be removed in the second version; however, it is matched with some random warning incorrectly. 

All the manual inspection is done by two code reviewers. The code reviewers first inspect the cascaded warnings by themselves and then compared and discussed the results with another code reviewer so that we report confident results.

\paragraph{Running experiments}
We ran all of our experiments on RedHat 20.4 Linux distribution on a virtual machine with 32 GB of memory and 32 cores available. We implemented our tools using LLVM-8.0, Python 3 and Bash scripts.

\subsection{Results for RQ1}

\subsubsection{Results for the ground-truth setting}

\begin{table}[htb]
	\centering
	\caption{\ben{Successful cascading for buggy-buggy versions}}
    \label{tbl:RQ1-buggy-buggy}
	\begin{tabular}{l|||c|c|c|c}\hline
   Benchmark         &Total  & Hydrogen & SCALe & GumTree \\\hline\hline
   ABM    & 20  & 20     & 19       & 20 \\\hline
   Toyota & 37  & 36     & 35       & 36 \\\hline
   Both    & 57  & 56     & 54       & 56 \\\hline
	\end{tabular}
\end{table}

\begin{table}[htb]
	\centering
	\caption{\ben{Successful cascading for buggy-fixed versions}}
	\label{tbl:RQ1-buggy-fixed}
	\begin{tabular}{l|||c|c|c|c}\hline
   Benchmark         &Total  & Hydrogen & SCALe & GumTree \\\hline\hline
   ABM    & 13  & 13     & 3        & 10 \\\hline
   Toyota & 26  & 26     & 5        & 23 \\\hline
   Both   & 39  & 39     & 8        & 33 \\\hline
	\end{tabular}
\end{table}

Tables~\ref{tbl:RQ1-buggy-buggy} and \ref{tbl:RQ1-buggy-fixed} show the results of each cascading tool for the benchmarks of ABM and Toyota. We leverage the ground truth bugs and patches in the benchmarks to confirm the successfulness of our warning cascading. Table \ref{tbl:RQ1-buggy-buggy} lists a total of 57 (20 from ABM and 37 from Toyota) pairs of programs in the two benchmarks.  Each pair of programs contains two buggy versions, where each version contains one bug, and the second buggy version is a refactored version of the first buggy version. The two versions contain the same bug. In Table \ref{tbl:RQ1-buggy-fixed}, under {\it Total}, we show that there are 39 (13 from ABM and 26 from Toyota) pairs of programs. Each pair of programs contains a buggy version and a fixed version. The buggy version contains one bug. The fixed version is the patched version for this bug. For both the cases, we focused on the static warnings generated for the buggy lines for this study (the fault locations are provided by the benchmarks) and determine if this warning is successfully cascaded.

In Tables \ref{tbl:RQ1-buggy-buggy} and~\ref{tbl:RQ1-buggy-fixed}, each cell in columns \textit{Hydrogen}, \textit{SCALe}, and \textit{GumTree} reported the number of targeted warnings that were successfully cascaded by the tool. Specifically, following in the criteria in Section 2.2, for Table \ref{tbl:RQ1-buggy-buggy}, we say a tool made a successful cascading if each warning of the buggy version in the first version is aligned with the warning of the same bug in the second version. For Table \ref{tbl:RQ1-buggy-fixed}, we say a tool made a successful cascading if each warning of the buggy line (first versions) did not find any match on a fixed version (second version). Here, the bug is fixed, and there is no warning for the same bug in the second version. Thus the warning in the first version should not match any other random warnings in the second version. 

As shown in Table~\ref{tbl:RQ1-buggy-buggy}, out of the 57 buggy-buggy program pairs, \texttt{Hydrogen} and \texttt{GumTree} cascaded 56 paired programs successfully, followed by \texttt{SCALe} at 54.  For buggy-fixed pairs, as shown in Table~\ref{tbl:RQ1-buggy-fixed}, out of the total 39 pairs of programs, \texttt{Hydrogen} was able to correctly cascade all 39 of them, followed by \texttt{GumTree} at 33 and \texttt{SCALe} at 8. The results based on the known ground truths show that \texttt{Hydrogen} outperformed the other baselines by successfully cascading 95 out of the total 96 pairs, followed by \texttt{GumTree} of 89 successful pairs and \texttt{SCALe} with 62 successful pairs. SCALe performed very poorly for cascading warnings for buggy-fixed pairs.





\subsubsection{Results for the real-world setting}

In this section, we compared the output generated from the three tools by cascading the warnings from real-world benchmarks and displayed the results using the Venn diagrams. We ran static analysis tools to generate the warnings for 12 pairs of programs. After obtaining the warnings, we did a preprocessing step before providing the warnings to the cascading tools, which included: 1) removing all the irrelevant warnings that have no effect on the execution of the programs, e.g., files from testing folders, obsolete library code, 2) aggregating warnings from different static analysis tools and removing all the duplicate warnings that are reported by different static analysis tools.  After the preprocessing step, the warnings (of the first version) were reduced from 19305 to 2113. These are the warnings we used for cascading.   

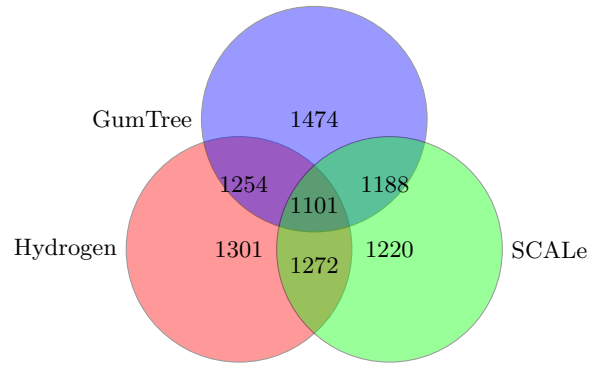
\begin{figure}
    \centering
    \begin{tikzpicture} [set/.style = {draw,
        circle,
        minimum size = 3cm,
        opacity = 0.4,
        text opacity = 1}]
     
        \node (A) [set,fill=red, label={180: Hydrogen}] {$1301$};
        \node (B) at (60:2cm) [set,fill = blue , label={-180: GumTree}] {$1474$};
        \node (C) at (0:2cm) [set, fill = green , label={0: SCALe}] {$1220$};
            
        \node at (barycentric cs:A=1,B=1) [left] {$1254$};
        \node at (barycentric cs:A=1,C=1) [below] {$1272$};
        \node at (barycentric cs:B=1,C=1) [right] {$1188$};
        \node at (barycentric cs:A=1,B=1,C=1) [] {$1101$};
     
    \end{tikzpicture}
    \caption{Number of warnings cascaded as {\it same "bugs"} (defined in Section 2.2)}
    \label{venn-matched}
\end{figure}

\begin{figure}
    \centering
    \begin{tikzpicture} [set/.style = {draw,
        circle,
        minimum size = 3cm,
        opacity = 0.4,
        text opacity = 1}]
     
        \node (A) [set,fill=red, label={180: Hydrogen}] {$812$};
        \node (B) at (60:2cm) [set,fill = blue , label={-180: GumTree}] {$639$};
        \node (C) at (0:2cm) [set, fill = green , label={0: SCALe}] {$893$};
         
        \node at (barycentric cs:A=1,B=1) [left] {$592$};
        \node at (barycentric cs:A=1,C=1) [below] {$764$};
        \node at (barycentric cs:B=1,C=1) [right] {$606$};
        \node at (barycentric cs:A=1,B=1,C=1) [] {$576$};
     
    \end{tikzpicture}
    \caption{Number of warnings cascaded as {\it "bug" fixes} (defined in Section 2.2)}
    \label{venn-unmatched}
\end{figure}
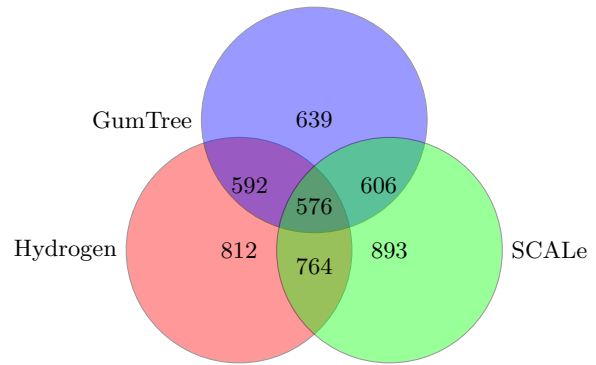

 In Figures \ref{venn-matched} and \ref{venn-unmatched}, the blue, red, and green circles represent GumTree, Hydrogen, and SCALe results respectively. Figure \ref{venn-matched} is a Venn diagram to show how many warnings are  cascaded to the same "bugs" between the two versions. Figure \ref{venn-unmatched} shows how many of warnings are cascaded as the {\it "bug" fix} between the two versions.

 The numbers located in the intersections of circles represent the shared cascading results across multiple tools. For example, 1101 on the center of Figure \ref{venn-matched} represents warnings that have the same cascading across the three tools. 1254 at the intersection of the blue and red circles represents warnings that have the same cascading between Hydrogen and GumTree, including the ones shared by the three tools. The number located in the circle alone (not in the intersection area) represents the number of warnings cascaded by that tool, including the ones shared with other tools. For example, 1301 in the center of the red circle in Figure \ref{venn-matched} means the total number of warnings cascaded by Hydrogen. Venn diagrams show that the three tools share a large part of warning cascading results, which brings in confidence that these cascading results are correct 
 

 

We randomly selected 12\% of total warnings from the results, covering each category of overlapped sets. The results of manual inspection are shown in Tables~ \ref{tbl:RQ2-matched} and \ref{tbl:RQ2-unmatched}.  The numbers presented in Tables 3 and 4 represent warnings of the first version that has been successfully cascaded. For example, in Table \ref{tbl:RQ2-matched} Row {\tt find} and Column {\it Total}, we show that we inspected a total of 33 warnings from the first version that are cascaded as same "bugs".  Hydrogen successfully cascaded all the warnings, SCALe successfully cascaded 31 and GumTree matched 32. As shown in Table~\ref{tbl:RQ2-matched}, out of the total 132 warnings that we inspected, \texttt{Hydrogen} was able to successfully cascade all 132 of them, followed by \texttt{GumTree} and \texttt{SCALe} with the same result at 99.

The results of "bug" fixes type of cascading are shown in Table~\ref{tbl:RQ2-unmatched}. Out of the 120 total warnings, \texttt{SCALe} cascaded 112 warnings correctly, followed by \texttt{GumTree} at 91 and \texttt{Hydrogen} at 85. It indicated that compared with other tools, Hydrogen has higher accuracy when detecting cascading bugs, and SCALe has more accuracy when determining if the "bug" is fixed. GumTree performed as a middle ground where have a better result than Hydrogen on cascading related to fixes and a better results than SCALe on cascading bugs. 

\begin{table}[htb]
	\centering
	\caption{\ben{Successful cascading of same "bugs" in real-world program pairs}}
	\label{tbl:RQ2-matched}
	\begin{tabular}{l|||c|c|c|c}\hline
   Benchmark         &Total  & Hydrogen & SCALe & GumTree \\\hline\hline
   find    & 33  & 33     & 31        & 32 \\\hline
   grep    & 22  & 22     & 2        & 20 \\\hline
   make    & 25  & 25     & 25        & 20 \\\hline
   coreutil & 52   & 52     & 41        & 27 \\\hline
   Total   & 132  & 132     & 99        & 99 \\\hline
	\end{tabular}
\end{table}

\begin{table}[htb]
	\centering
	\caption{\ben{Successful cascading of "bug" fixes in real-world program pairs}}
	\label{tbl:RQ2-unmatched}
	\begin{tabular}{l|||c|c|c|c}\hline
   Benchmark         &Total  & Hydrogen & SCALe & GumTree \\\hline\hline
   find    & 21  & 20     & 24        & 1 \\\hline
   grep    & 24  & 24     & 21        & 20 \\\hline
   make    & 10  & 10     & 10        & 10 \\\hline
   coreutil & 65  & 31     & 57        & 60 \\\hline
   Total   & 120  & 85     & 112        & 91 \\\hline
	\end{tabular}
\end{table}




\subsection{Results for RQ2}
To answer RQ2, we further analyzed and grouped the warning cascading results from the three tools into the following categories:
\begin{enumerate}
    \item GumTree failed to cascade
    \item SCALe failed to cascade
    \item SCALe and GumTree failed to cascade
    \item Hydrogen failed to cascade
\end{enumerate}


\subsubsection{GumTree failed to cascade}
GumTree can fail for two reasons. The first reason is that GumTree cannot process macros in the programs. In the presence of macros, the ASTs sometimes are parsed incorrectly and do not match the source code. This prevents GumTree from matching the warnings correctly.
The second reason is that GumTree used a heuristic algorithm to map the AST nodes based on their syntax, regardless of their semantic meaning. This approach can lead to incorrect mapping of the AST nodes across versions, causing warning cascading to fail.

\begin{figure}
    \centering
    \begin{lstlisting}[language=C,
    frame=none,
    basicstyle=\footnotesize\ttfamily,]
static char **
construct_command_argv_internal
(char *line, char **restp, char *shell,
char *shellflags, char *ifs, int flags,
char **batch_filename_ptr)
{
    ...
    #ifdef _AMIGA
    <@\textcolor{green}{+ if(one\_shell)}@>
    <@\textcolor{green}{+ \{}@>
    <@\textcolor{green}{+ \#if defined \_\_MSDOS\_\_ || defined (\_\_EMX\_\_)}@>
    <@\textcolor{green}{+ if (unixy\_shell)}@>
    <@\textcolor{green}{+ \#else}@>
    <@\textcolor{green}{+ if (is\_bourne\_compatible\_shell(shell))}@>
    <@\textcolor{green}{+ \#endif}@>
    <@\textcolor{green}{...}@>
    
    ap = new_line;
    memcpy (ap, shell, shell_len);
    ap += shell_len;
    *(ap++) = ' ';
    memcpy (ap, shellflags, sflags_len);
    ap += sflags_len;
    *(ap++) = ' ';
    <@\textcolor{blue}{command\_ptr = ap;}@>
    ...
    #endif _AMIGA
}
    \end{lstlisting}
    \caption{GumTree failed due to macro}
    \label{gum_fail1}
\end{figure}

Figure~\ref{gum_fail1} shows an example that only GumTree made the wrong cascading among the three tools. In this example, a section of code is surrounded by a conditional compilation using the macro \texttt{\_AMIGA}. The presence of such a macro, caused the AST to parse the information in the incorrect way. In the AST diff, this region of code is considered as deleted in the second version, which fails to match the rest of the AST nodes in the two versions. SCALe and Hydrogen both can handle such a case and made a correct cascading.


\begin{figure}[ht]
  \centering
  \includegraphics[width=1.0 \linewidth]{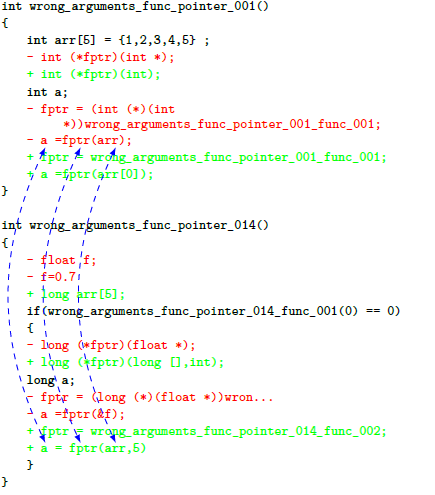}
  \caption{GumTree failed to cascade {\tt Toyota} warnings because AST diff algorithm cannot align ASTs of the two versions correctly}
  \label{fig:div_con}
\end{figure}
GumTree makes mistakes also because of its syntactic diff algorithm. In Figure \ref{cvsex2}, we showed diffs of two functions: \texttt{wrong{\textunderscore}001} (abbreviated) and \texttt{wrong{\textunderscore}014}. GumTree AST diff algorithm incorrectly matched the first version of \texttt{wrong{\textunderscore}001} to the second version of \texttt{wrong{\textunderscore}014} (an irrelevant function), instead of  the second version of \texttt{wrong{\textunderscore}001}. The blue arrow shows the nodes which GumTree considers to be matched between the two versions.
The nodes \texttt{a}, \texttt{fptr}, and \texttt{arr} in function \texttt{wrong{\textunderscore}001} were mapped to identify nodes in the irrelevant function \texttt{wrong{\textunderscore}014}, but they should have been mapped to the version 2 of \texttt{wrong{\textunderscore}001}.
This incorrect mapping caused the warning on line 8 to be incorrectly cascaded to line 26 (version 2 of \texttt{wrong{\textunderscore}014}) instead of line 10 (version 2 of \texttt{wrong{\textunderscore}001}).


\subsubsection{SCALe failed to cascade}
SCALe can generate two types of errors when performing warning cascading. 
First, two snippets of code may match textually but a change in referenced elements, e.g., a change in the called function, can cause different execution behaviors.
SCALe can incorrectly match the warnings. Second, some textual differences have no impact on program behaviors related to the warnings, but SCALe would falsely report the warnings cannot match. In the following, we further provide two examples to demonstrate the weakness of SCALE.

\begin{figure}
    \centering
    \begin{lstlisting}[language=C,
    numbers=right,
    frame=none,
    xleftmargin=0.5cm,
    xrightmargin=1.5cm,
    basicstyle=\footnotesize\ttfamily,]
static size_t
cwexec (kwset_t kws, char const *text,
size_t len, struct kwsmatch *kwsmatch)
{
    struct kwset const *kwset;
        ...
    <@\textcolor{red}{- unsigned char c;}@>
    <@\textcolor{red}{- unsigned char const *delta;}@>
    <@\textcolor{red}{- int d;}@>
    <@\textcolor{red}{- char const *end, *qlim;}@>
    <@\textcolor{red}{- struct tree const *tree;}@>
    <@\textcolor{red}{- char const *trans;}@>
    <@\textcolor{green}{+ register unsigned char c;}@>
    <@\textcolor{green}{+ register unsigned char const *delta;}@>
    <@\textcolor{green}{+ register int d;}@>
    <@\textcolor{green}{+ register char const *end, *qlim;}@>
    <@\textcolor{green}{+ register struct tree const *tree;}@>
    <@\textcolor{green}{+ register char const *trans;}@>
    ...
    kwset = (struct kwset *) kws;
    if (len < kwset->mind)
        <@\bf{\color{blue} return -1;}@>
    ...
}
    \end{lstlisting}
    \caption{SCALe failed to cascade {\tt grep} warnings}
    \label{img:cvsex}
\end{figure}

Figure~\ref{img:cvsex} shows the diff between versions \texttt{401d8194} (shown in red) and \texttt{54d55bba} (green) of \texttt{kwset.c} in \texttt{Grep}. The static analysis tool reports a warning at line 22 (shown in blue) for both the versions as \textit{-1 is coerced from int to unsigned long}. SCALe, however, reports the lines 7-22 as being modified instead of just the lines 7-18. Due to the line containing the warning (line 22) being misclassified as modified, SCALe fails to cascade this warning. However, this  warning should have been matched because (1) line 22 was not modified between the versions and (2) adding the keywords "register" at lines~13--19 should not change the semantics related to this type of bug.

\begin{figure}
    \centering
    \begin{lstlisting}[language=C,
    frame=none,
    basicstyle=\footnotesize\ttfamily,]
static void
bytes_split (uintmax_t n_bytes, char *buf, 
size_t bufsize)
{
    size_t n_read;
    bool new_file_flag = true;
    size_t to_read;
    uintmax_t to_write = n_bytes;
    <@\textcolor{blue}{char *bp\_out;}@>
    <@\textcolor{green}{+ uintmax\_t opened = 0;}@>
    ...
}
    \end{lstlisting}
    \caption{SCALe fails to casade}
    \label{Scale_fail2}
\end{figure}
Figure~\ref{Scale_fail2} shows an example that only SCALe made the wrong match cascading among the three tools. Similar to the example shown in Figure \ref{img:cvsex}, this program has a newly added line below the target line that we aim to cascade. Due to this change, the Unix diff reported that the target line has changed in the new version, which caused SCALe to fail to match warnings where AST and CFG-based diff tools can cascade successfully.


\subsubsection{SCALe and GumTree failed to cascade}
\begin{figure}
    \centering
    \begin{lstlisting}[language=C,
    frame=none,
    basicstyle=\footnotesize\ttfamily,]
    
<@\textcolor{green}{+ \# MACRO\_BEGIN}@>
char
human_readable (...
<@\textcolor{red}{- uintmax\_t from\_block\_size, uintmax\_t to\_block\_size)}@>
    <@\textcolor{green}{+ uintmax\_t from\_block\_size, uintmax\_t to\_block\_size)}@>
{
    ...
    <@\textcolor{blue}{unsigned int base = human\_base\_1024 ? 1024 : 100;}@>
    ...
}
<@\textcolor{green}{+ \# MACRO\_END}@>
    \end{lstlisting}
    \caption{GumTree and SCALe fail to cascade}
    \label{Gum_Scale_fail}
\end{figure}
Figure \ref{Gum_Scale_fail} illustrates a scenario in which both GumTree and SCALe made incorrect warning cascading. In the case of GumTree, the failure is due to a large macro that surrounds the warning location on the second version. This macro makes it difficult for GumTree to parse the code into an AST, which results in a failure to match the warnings. In the case of SCALe, the failure is caused by a difference in the text located directly above the warning statement. This difference affects the results of the Unix diff tool but does not change the semantics of code. On the other hand, Hydrogen will not be affected by such changes because 1) it successfully parses the code within macro and built it into the MVICFG; 2) it uses control flow graphs to perform diff. It can confirm that this change does not affect program control flow and semantics, and thus will mark the blue statement as the matched warnings in MVICFG (See algorithm \ref{alg:CascadeWarning}).

The main reason Hydrogen fails in some cases is that it used LLVM to compile the program, and some code that cannot be handled by LLVM is excluded from MVICFG. For example, {\tt if} statements that do not have brackets, a statement expanding across multiple lines (it covers 46.8\%  undetected cascading for Hydrogen),  the internal function used for library and some conditional compilation code is not covered by this build.

Hydrogen used textual diff tool to build MVICFG and sometimes has the disadvantage similar to SCALe. In Section 2.1, Figure~\ref{img:chg3} shows a snippet of code where function {\tt fprintf} is refactored to {\tt checked\_fprintf}. In this example, the text has changed between two versions of a related statement. However, the functionality still remains the same. Thus the warnings should be cascaded as the same "bug". Hydrogen fails to cascade this warning correctly because the target line has been marked as modified (unmatched node in MVICFG). If the textual statement wasn't changed, Hydrogen could possibly make a correct detection by using algorithm \ref{alg:CheckPaths}. However, since the function names in the statements have been changed, Hydrogen is not able to handle it. GumTree is able to make a correct cascading by leveraging the AST structures. 

\section{Threats to Validity}
To address the external threats to validity, we applied 12 pairs of C real-world projects as well as 96 pairs of benchmark programs where we know ground truth. We applied a set of static analysis tools often used by CERT to make sure we generated all types of practical static warnings. The benchmarks also had varying commits between versions to ensure heterogeneous diffs.


To address the internal threat to validity, we first inspected the output of three tools to make sure the implementations of the warning cascading algorithms are correct. We inspected 100\% of cascaded warnings from the ground truth and 12\% for real-word programs across by two code reviewers to confirm our findings.

\section{Related Work}


There have been many works that focus on matching and prioritizing warnings or faults between multiple versions of a program~\cite{Fischer2003PopulatingaRelease,Logozzo2014VerificationModuloVersions, Palix2015Improvingpatterntracking, Spacco2006Trackingdefectwarnings, Kim2007Whichwarningsshould, Kim2007Prioritizingwarningcategories, Boogerd2009Evaluatingtherelation,  Chimdyalwar2011Effectivefalsepositive, Avgustinov2015Trackingstaticanalysis} . \cite{Fischer2003PopulatingaRelease, Chimdyalwar2011Effectivefalsepositive, Logozzo2014VerificationModuloVersions,  Palix2015Improvingpatterntracking} uses GNU diff, AST, and Verification Modulo Versions to provide matching, while \cite{Kim2007Whichwarningsshould, Kim2007Prioritizingwarningcategories} use source control revisions to prioritize static warnings.
To the best of our knowledge, there has been no study on tools of warning cascading.


Spacco et al. \cite{Spacco2006Trackingdefectwarnings} developed two methods to match warnings in the static analysis tool FindBugs at the line granularity level. 
The first approach, `pairing', matches warnings based on their source code location. First, it identifies exact matches of package, class, and method name. Then, for those warnings that do not match exactly, the approach uses progressively `fuzzier' criteria. The second approach is called `warning signatures'. This approach transforms each warning into a string format that includes information about the warning, and then matches string-formatted warnings with the same MD5 code. Both the `pairing' and `warning signature' methods perform best when using a single static analyzer tool and use textual diff to identity places to do the cascading. 

Logozzo et al. \cite{Logozzo2014VerificationModuloVersions} present a solution to the common problem of verifying software with multiple versions. To tackle this challenge, the authors introduce a novel verification framework called Verification Modulo Versions (VMV), which is specifically designed to enhance the efficiency and effectiveness of software verification for multi-version systems. While this work does verification and relies on their own framework, our work tackles cascading warnings generated by off-the static analyzers and compares the usefulness and efficiency of different cascading methods independently of the specific analyzer.

Palix et al. \cite{Palix2015Improvingpatterntracking} build an AST based on code changes to improve tracking changes similar to GumTree\cite{Falleri2014Gumtree}.
However, their work focuses on tracking changes between multiple versions, while our work focuses on studying how different change-tracking methodologies affect warning cascading.


Finally, there are many methods to track changes~\cite{Yang1991IdentifyingSyntacticDifferences, Pawlik2011RTED, Fluri2007Changedistilling, Falleri2014Gumtree,ASTChange2017,Frick2020UnderstandingSC} which can be roughly separated into textual, syntactic, and semantic methods.
None of them directly deals with the problem of matching warnings between versions, but some of them are used in other works about matching warnings. Here, we discuss the state of the art for each category.

Syntactic methods such as GumTree \cite{Falleri2014Gumtree} work at the granularity of ASTs, which reflects the source code structure and hence can be more precise than textual diff. AST helps in avoiding common pitfalls of textual based diff like missing refactoring based changes and spacing issues.  Yang et al. \cite{Yang1991IdentifyingSyntacticDifferences} developed a syntactic-based comparing method for dynamic programming languages like scheme. Similar to GumTree, Fluri et al. \cite{Fluri2007Changedistilling} also proposed an AST based approach using a tree-differencing algorithm to detect source code changes. We choose GumTree because it is recent, open-source, and has been widely used by other works. Huang et al. \cite{CLDiff2018} present an approach called ClDiff to linked code differences with the aim of simplifing code review. While these works aim to improve tracking changes between versions, our work focuses on studying the impact of using different tracking methods for cascading warnings.

\section{Conclusions and Future Work}
Cascading static warnings is a practical but challenging problem. This paper applied three tools to explore their pros and cons of addressing this problem. We found that SCALe, the textual diff based tool, fails when there are textual changes but not semantics changes related to the bugs. It also fails when the referred calls or global variables have changed outside the current functions. GumTree has the weakness of not being able to handle macros, and the AST tree matching algorithm faces some failures due to its heuristic nature. Hydrogen relied on LLVM and cannot process all the code in the repositories due to the requirement of building the project. It used textual diff tool to build MVICFG and sometimes has the  disadvantage similar to SCALe. In the future, we plan to integrate more static analysis tools like {\tt CodeSonar}. Such tools produce paths as static warnings, and we envision that CFG based diffs can have greater advantages. 

\bibliographystyle{IEEEtran}
\bibliography{ref}

\end{document}